\documentclass{elsart}
\usepackage[dvips]{graphics}  
\begin{document}
\runauthor{Cicero, Caesar and Vergil}

\begin{frontmatter}

\title{Structure of Rule Table and Phase Diagram of One Dimensional
Cellular Automata}

\author{Sunao Sakai and Megumi Kanno}

\address{Faculty of Education, Yamagata University, Yamagata,
990-8560, Japan}\thanks{E-mail: sakai@e.yamagata-u.ac.jp}

\begin{abstract}
 In addition to the $\lambda$ parameter, we have found another
parameter 
which characterize the class III, class II and class IV patterns 
more quantitatively.
It explains why the different classes of patterns coexist
at the same $\lambda$. 
With this parameter, the 
phase diagram for an one dimensional cellular automata is
obtained. 
Our result
explains why the edge of chaos(class IV)
is scattered rather wide range in $\lambda$ around $0.5$, 
and presents an effective way to 
control the pattern classes.

\end{abstract}

\maketitle

\end{frontmatter}

\section{Introduction}

The study of the cellular automata(CA) has been one of the most 
important fields for the understanding of the
complex systems. And many efforts have been carried
out for the understanding of it.  
Various patterns had been generated
by choosing the transition rules, and the patterns 
have been classified by Wolfram roughly into four
classes\cite{wolfram}. Langton has introduced parameter $\lambda$ and
showed that as $\lambda$ is increased the pattern
changes from class I(homogeneous),class II(periodic) to class
III(chaotic). And at some cases, the class IV(edge of chaos) is
realized between the two\cite{langton,langton2}.\\

The $\lambda$ parameter could classify the qualitative behavior of the 
CA, but it is not enough\cite{mitchell}. It is well known that at the
same $\lambda$ the time evolution of the configuration shows periodic
pattern(class II), chaotic one(class III) or edge of the chaos
pattern(class IV) depending on the initial seed
of the random number generator.
The reason or mechanism for it has not been known yet;  
we have no way to control the pattern classes at fixed $\lambda$.
And the transition of periodic to
chaotic pattern is taken place at rather wide range of $\lambda$.
Therefore it is a natural anticipation
that more parameters are necessary to get the quantitative
understanding of the pattern classes of the CA. In
this article
we report another parameter which classifies the pattern classes more
quantitatively.
If the capacity to support computation is realized at the
edge of the chaos\cite{langton,mitchell}, we have acquired a 
method to reach there more effectively.

In the section 2, we will briefly summarize our notations and explain
the key discovery which leads us to the understanding of the structure
of the rule table. It strongly indicates that the rules which destroy
the schema of the quiescent state play an important role to
distinguish the pattern classes.

In the section 3, guided by the discovery of the section 2, we
classify the rule table according to the
destruction and construction of the schema of the quiescent states.
The rule which destroys the longest schema of the
quiescent state plays strongest effects for the patterns. 
Then we introduce a new parameter force for each rules which destroy the
schema.
With this 
parameter, we determine the phase diagram in the $\lambda$ and
the force plane. 
It clearly explains why at fixed $\lambda$ the different
patterns are generated and we could control the pattern classes
more quantitatively.

Section 4 is devoted to the discussions and the conclusions, where 
effects of the structure of the rule table 
will further be discussed.

\section{Discovery of the Key Rules}

\subsection{Summary of the Notation}
In order to make the argument concrete, in this article we focus on 
the one dimensional
cellular automata with 5-neighbors and 4-states.  
But the qualitative results are independent of the details of the 
models. This point will be discussed in section 4.
We follow the
definitions of Langton\cite{langton} but for the self consistentness we
will briefly summarize them.

In our study, the sites consist of
150 cells with the periodic boundary condition, which are denoted as
$s(t,i)$. 
The $t$ represents the time step which take integer value,
and 
the $i$ is the position of site which range from $0$ to
$149$. 
The $s(t,i)$ takes the value $0,1,2$ and $3$, and the state $0$ is taken
to be quiescent state\cite{langton}.  The set of the states $s(t,i)$
at fixed $t$ is called the configuration at time $t$.  

The configuration at time $t+1$ 
is determined by that of time $t$ by the following local relation,

\begin{equation}
s(t+1,i) = T(s(t,i-2),s(t,i-1),s(t,i),s(t,i+1),s(t,i+2)) \label{table}
\end{equation}

The set of the mappings 
\begin{equation}
T(\mu,\nu,\kappa,\rho,\sigma)=\eta,(\mu,\nu,etc= 0,1,2,3) 
\label{r_table}
\end{equation}
is called
rule table.  The rule table consists of $4^5$ mappings,
which are selected from the totally $4^{1024}$ elements.

The $\lambda$ parameter is defined as follows\cite{langton},
\begin{equation}
\lambda=\frac{N_{h}}{1024} \label{lambda}
\end{equation}
where $N_{h}$ is the number that $\eta$ in Eq.\ref{r_table} is not
equal to $0$. In other word 
the $\lambda$ is a probability that the rules do not select the
quiescent state in next time step. In the following we set the
rule tables randomly according to the probability
$\lambda$\cite{langton}, 
and the initial configurations are also set randomly.

The time history of the configurations 
is called pattern. The patterns are classified by
Wolfram\cite{wolfram} roughly into four classes. It has been known
that as the
$\lambda$ increases the most frequently generated 
patterns change from homogeneous(class I),
periodic(class II) to chaotic(class III)  
and at the region between class II and class III, the  edge of
chaos is located(class IV). 

\subsection{Experiments at $\lambda=0.44$ and Discovery of the Key Rule}

In order to find the reason why the different patterns are generated
at the same $\lambda$, 
we had started to collect a lot of rule tables with 
different pattern classes. And by studying them, we had tried to
find 
the differences between rule tables.
We had fixed $N_{h}=455$, which correspond to $\lambda=0.44$.
We had chosen this $\lambda$
because at this point the chaotic, edge of chaos and periodic patterns
are generated in the similar ratio. 
By changing the random number, we have
gathered a few tens of the rule tables and classified them
into chaotic, edge of chaos and periodic ones.

First we studied whether the pattern classes are sensitive to the
initial configurations or not.
We fixed the rule table and changed the initial configurations.
The details of the patterns depend on the initial configurations but
the pattern classes are not changed\cite{wolfram}.
Then the difference of the pattern classes is due to that of the
rule tables and 
our target is addressed to the study of them.

After some trial and error, we have
found the fact that for rule tables which generate the edge of the
chaos and periodic patterns, the transition table $T(0,0,0,0,0)$ was
almost always equal to $0$, while for that of the chaotic pattern, it
was not equal to $0$ in majority of cases.
For about 10 rule tables of each 
pattern classes, we get following results
for the probability that the rule tables have $T(0,0,0,0,0)=0$.
\begin{equation}
Prob(T(0,0,0,0,0) = 0) \sim 1 \label{perio}
\end{equation} 
for periodic plus edge of chaos rule tables, \\
\begin{equation}
Prob(T(0,0,0,0,0) =0 )\sim 0.2 \label{chaos} 
\end{equation}
for chaotic rule tables. 

In this article, we call the pattern as edge of chaos when its
transient length\cite{langton} is longer
than $3000$ time steps.

These difference given by Eq.\ref{perio} and Eq.\ref{chaos} has
lead us the following idea.
Let us think of the schemas of quiescent state $0$ with length longer
than five. 
If the rule $T(0,0,0,0,0)=h,(h \neq 0)$ is included in the rule
table, they are necessarily  broken
\footnote{Similar idea has been
pointed out by Wolfram and Suzudo with the
arguments of the unbounded growth\cite{wolfram} and
expandability\cite{suzudo}.}.
Then it pushes the pattern to that of chaos.

\begin{figure}
\begin{center}
\scalebox{0.55}{ { \includegraphics{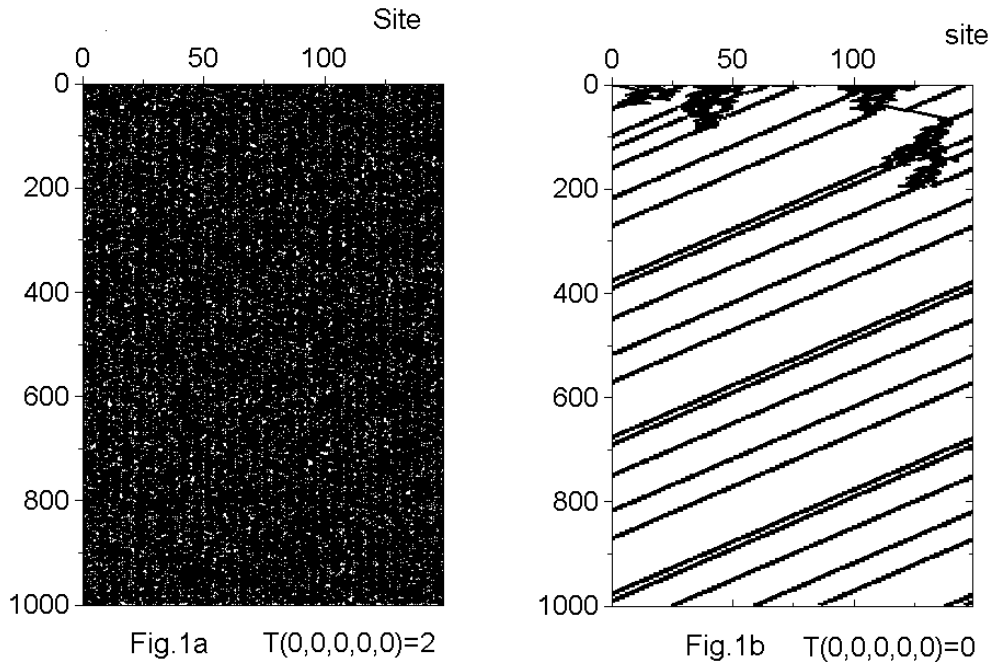} } }
\caption{ An example of the replacements experiment at $\lambda=0.44$.
The quiescent state is shown with white points and the other states, 
with black one. The rule $T(0,0,0,0,0)=2$ in the Fig.1a is replaced by
$T(0,0,0,0,0)=0$ in the Fig.1b, and at the same time a replacement
$T(\mu,\nu,k,\rho,\sigma)=0,(k \neq 0) \rightarrow 
T(\mu,\nu,k,\rho,\sigma)=l,(l \neq0)$ 
has been carried out to keep the $\lambda$ unchanged. 
\label{pattern_change} }
\end{center}
\end{figure}

If our idea is correct, we could artificially change the rule tables
of the chaos to that of periodic by replacing $T(0,0,0,0,0)=h$ to
$T(0,0,0,0,0)=0$.
We have tried the replacements for about 10 rule tables.
In all cases we have succeeded to change the rule tables from 
chaotic to periodic or  
periodic to chaotic by a replacements of the $T(0,0,0,0,0)$. 
An example is shown in the Figure  \ref{pattern_change}.
The change in the pattern is dramatical but the differences in the 
rule table are only two, which has been quite exciting for us.
This discovery was a key hint to lead us to the hypothesis
that the rules which breaks the schema of the quiescent state will
determine the pattern classes.

\section{Structure of Rule Table and Phase Diagram}

\subsection{Classification  of the Rule Table} 

In order to test the hypothesis of the previous section,
we group the rule table into the four types according to the
operation on the schema of 
quiescent state.  

In the following, Greek characters in the rule table
represent $0,1,2$ and $3$ while Roman ones $1,2$ and $3$.

Type 1: $T(\mu,\nu,0,\rho,\sigma)=h$.\\
The rules in this type destroy the schemas of quiescent state.

Type 2: $T(\mu,\nu,0,\rho,\sigma)=0$. \\
The rules of this type conserve them.\\

Type 3: $T(\mu,\nu,k,\rho,\sigma)=0$.\\
The rules of this type develop the schema.\\

Type 4: $T(\mu,\nu,k,\rho,\sigma)=l$. \\
The rules in this type do not affect the schema in next time step.

The sum of the number of the type 1 and type 2 rules is 256, while that of
type 3 and type 4 rules is 768. How many numbers of each type of rules are 
included
in the rule table is determined randomly according to the probability
 $\lambda$. 

The type 1 rules are further classified into five groups according to
the length of the schema which breaks. They are
shown in the Table \ref{destruction}. \\
  \\

\renewcommand{\arraystretch}{0.9}
\begin{table}[h]
\begin{center}
\begin{tabular}{|c|c|c|c|c|c|c|}
     \hline
     \multicolumn{1}{|c|}{group} &
     \multicolumn{1}{|c|}{Total Number} &
     \multicolumn{1}{|c|}{Name}&
     \multicolumn{1}{|c|}{Replacement}\\
     \hline
        $T(0,0,0,0,0)=h$  &1  &D5 &RP5\\
     \hline
        $T(0,0,0,0,i)=h$  &3  &D4  &RP4\\
        $T(i,0,0,0,0)=h$  &3  &    &    \\
     \hline
        $T(0,0,0,i,\sigma)=h$ &12 &   &    \\
        $T(i,0,0,0,m)=h$      &9  &D3 &RP3\\
        $T(\mu,j,0,0,0)=h$    &12 &   &    \\
     \hline
        $T(\mu,j,0,0,m)=h$    &36 &D2 &RP2   \\
        $T(i,0,0,l,\sigma)=h$ &36 &   &       \\
     \hline
        $T(\mu,j,0,l,\sigma)=h$ &144 &D1 &RP1 \\
     \hline
\end{tabular}  
\end{center}  
\vspace{0.5cm}
\caption{ The classification of the rules in type 1.
 \label{destruction} }
\vspace{0.5cm}
\end{table}

Our hypothesis presented at the end of the
previous section is expressed more quantitatively as follows;
what groups of rules and how many of them shown in the table
\ref{destruction} are
included in the rule table will determine whether the
resulting patterns become chaotic, edge of chaos, or periodic.

In order to test this hypothesis we have artificially replaced
some of the rules 
in the table \ref{destruction} while keeping
the $\lambda$ fixed.
For the D5, it is a set of the replacements given by the following
equations, which are already carried out in obtaining 
Fig.\ref{pattern_change}.

\begin{equation}
\begin{array}{ll}
T(0,0,0,0,0)=h \rightarrow T(0,0,0,0,0)=0 \\ 
T(\mu,\nu,k,\rho,\sigma)=0 \rightarrow T(\mu,\nu,k,\rho,\sigma)=l\\
\label{toperio} 
\end{array}
\end{equation}
where except for $h$, the $\mu$, $\nu$, $\rho$, $\sigma$, $k$ and $l$ 
are selected randomly.

Similarly the replacements are generalized for  D4 to D1 in the table
\ref{destruction}, which are named by RP5 to RP1 there. 
They change the rules in type 1 to
that of the type 2 and push the rule table toward
periodic one.

The reverse replacements for D5 are,
\begin{equation}
\begin{array}{ll}
T(0,0,0,0,0)=0 \rightarrow T(0,0,0,0,0)=h \\ 
T(\mu,\nu,k,\rho,\sigma)=l \rightarrow T(\mu,\nu,k,\rho,\sigma)=0\\
\label{tochaos} 
\end{array}
\end{equation} 
which push the rule table toward chaotic one. In this case $h$ is 
selected randomly, while $l$ is fixed in rule table.
Similarly we define the replacements for D4 to D1, which will be called 
RC5 to RC1 in the followings.

We have made replacements RP5 to RP1
if the initial rule table generates chaotic pattern and RC5 to RC1 if
it belongs to the periodic rules.
These replacement experiments are carried out for $\lambda=0.6$,$0.5$ 
and $0.4$. 

When the replacements include RP5, a few additional
replacements RP4s together with D5  make the chaotic pattern to
periodic ones and
converse is true for the RC5.
The numbers of the replacements needed to change the pattern are alway
less than the $1\%$ of the $1024$ elements of the rule tables.  

When the replacement does not include RP5 or RC5 the number of the
replacement needed to change the pattern classes increases. 
This means that
the effects to push the rule table toward periodic are different for
the replacements in 
table \ref{destruction}.  From a few tens of the replacements
experiments the following order for the effects are observed.
\begin{equation}
RP5 > RP4 > RP3 > RP2 > RP1  \label{order_force}
\end{equation} 

\subsection{Phase Diagram in Term of Force of Type 1 Rules}

In order to make the relation given by Eq. \ref{order_force} more 
quantitative
we introduce a new parameter force for type 1 rules.

We assume that each rule in the table
\ref{destruction} have their own force to push the rule tables toward the
chaos.
The force may be a complicated function of the numbers of type 1 to
type 4 rules, or it may even depend on the detailed contents of the
1024 rules. 
But as a first approximation, we take that it depends only on
numbers 
of the rules D5, D4, D3, D2 and D1 in table \ref{destruction}, 
which are denoted as $N_5$, $N_4$, $N_3$, $N_2$ and $N_1$ respectively.

Second we assume that the forces are represented by the simple
weighted sum  of the
number $N_5$ to $N_1$. The weight represents the strength of the force
and expressed by the coefficients $c_{5}$ to $  c_{1}$. 

In the study of the replacement of the rule tables, we found that the  
D1 has quite small force to change the rule table. In some cases the
replacement RP1 pushes the rules to chaotic one. 
In these cases it is close to the
random walk in the rule table space and the determination of $c_{1}$ 
needs much statistics\footnote{
The schemas of state $0$
with length 1 are easily created by the type 3 rules by one time step.
Then the force of D1 rule may easily be compensated by them. This may 
be a reason that 
the replacement RP1 some times looks like random walk.\\
}. 
In the following we neglect the 
contribution from D1 rule and set $c_{1}=0$ as a first approximation. 
Then the force is given as follows
\footnote{
This equation is interpreted in the following way.
The total number of the D5 rule is rather large that the fluctuations
from them are small. We replace them as the average background force
and measure the force from this background.}.

\begin{equation}
f(N_5,N_4,N_3,N_2)= c_5 N_5+c_4 N_4 + c_3 N_3 + c_2 N_2
\label{force_sum}
\end{equation} 

The ansatz given in Eq. \ref{force_sum} is a first step to 
the more detailed study of the structure of rule tables of the CA. 
The fine forces
of D2 or even by the D3 may depend on the fine structure of
the type 3 and type 4 rules. In this article these details 
are neglected.

By artificially carrying out the replacements of RP5 to RP2 or RC5 to RC2
we look for the critical combinations of the $N_5$ to $N_2$,
where the change of the patterns is observed.
We have collected  about 60 to 85 sets of critical combinations for each
$\lambda$. 
We assume that the transition occurs when the force given by
Eq.\ref{force_sum} exceeds at some fixed threshold value. The threshold
value is not known a priori and it may depend on the individual rule
table, but as a first approximation, we assume that 
it is common for a given $\lambda$. 
With this condition, 
we proceed to determine the coefficients.

However in this condition the scale for the force is arbitrary. We measure it
in the unit where force from D5 is equal to one, which correspond to 
divide the
force in Eq. \ref{force_sum} by $c_5$ and express it as the ratio
$r_4$, $r_3$ and $r_2$, where $r_4=\frac{c_4}{c_5}$, 
similarly for $r_3$ and $r_2$.
Then our problem is to find the $r_4$, $r_3$ and $r_2$, which minimize the
following quantity, 

\begin{equation}
s(r_4,r_3,r_2)= 
\frac{1}{c_{5}^{2}}\sum_{i,j}(f^i(N_5^i,N_4^i,N_3^i,N_2^i)
            - f^j(N_5^j,N_4^j,N_3^j,N_2^j))^{2}
\end{equation}

The results for $r_4$, $r_3$ and $r_2$
are given in the table \ref{relative_force}, 
where the errors are estimated by jackknife method.
They satisfy the
order given by the Eq.\ref{order_force}. This shows a
hierarchy in the forces of the rules in table \ref{destruction}.

\renewcommand{\arraystretch}{0.9}
\begin{table}[t]
\begin{center}
\begin{tabular}{|c|c|c|c|c|c|c|}
     \hline
     \multicolumn{1}{|c|}{$\lambda$} &
     \multicolumn{1}{|c|}{  } &
     \multicolumn{1}{|c|}{Relative Strength} &
     \multicolumn{1}{|c|}{Error}\\
     \hline
&                 $r_{4}$  &0.1126 &0.0034\\
$\lambda=0.4$&    $r_{3}$  &0.0418 &0.0013\\ 
&                 $r_{2}$  &0.0109 &0.0053\\
 \hline
&                 $r_{4}$  &0.1230 &0.0046 \\
$\lambda=0.5$&    $r_{3}$  &0.0432 &0.0014 \\
&                 $r_{2}$  &0.0119 &0.0012 \\
 \hline
&                 $r_{4}$  &0.1138 &0.0026 \\
$\lambda=0.6$&    $r_{3}$  &0.0275 &0.0034 \\
&                 $r_{2}$  &0.0201 &0.0012 \\
 \hline
\end{tabular}  
\end{center}  
\vspace{0.5cm}
\caption{The relative strength of forces for the groups of rules  
given by the table 
\ref{destruction}. \label{relative_force}
 }
\vspace{0.5cm}
\end{table}

The results in table \ref{relative_force} are natural in the 
following sense.
If six D4 rules are present for the 
rule table, then the presence of D5 has no effects, because the length 
5 schema of quiescent state could not be made. They have roughly 
similar effects
as one D5 rule, then the strength of the D4 will approximately be 
equal to $1/6$ of that
of D5. Similarly the strength of the D3 and D2 will be about $1/33$ and
$1/72$ respectively. The results given in the table
\ref{relative_force} are not very different from these naive
expectations.

\begin{figure}
\begin{center}
\scalebox{0.60}{ { \includegraphics{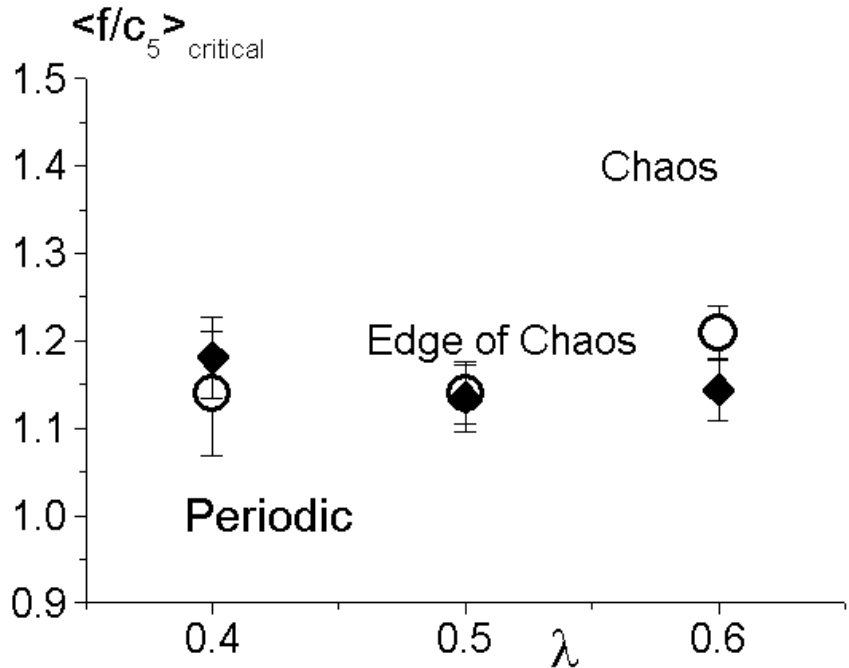} } }
\caption{Phase diagram of the 5-Neighbor 4-State cellular
Automata. The open circle is the $f_{critical}/c_5$ from the relative
strength of the table \ref{relative_force}, while the filled diamond
show the $f_{critical}/c_5$ from the naive expectations 
discussed in the text.
We should like to notice that force is divided by $c_5$, 
which may depends on $\lambda$. 
\label{phase_diagram} }
\end{center}
\end{figure}

With these results for the coefficients of Eq.\ref{force_sum}, we
determine the transition points by means of the force. We call it
as critical force($f_{critical}$). 
The results are shown in the Fig.\ref{phase_diagram}.
In the Fig.\ref{phase_diagram}, we have added 
$f_{critical}/c_5$ calculated from the 
naive expectations for the relative strengths. In this case too,
errors are estimated by the jackknife method.
The difference between them are not large. 

This figure distinguishes the pattern classes more quantitatively,  
and visualizes why at the fixed $\lambda$ the three patterns 
chaos, edge of the chaos and periodic
coexist.  

The impressive point is that the transition occurs about $f/c_5 \sim
1.14 - 1.2$. This means that in many cases the replacements of D5 will
change the pattern classes, which is consistent with
the experiments at section 2.

We should like to notice that the phase diagram shown in Fig.
\ref{phase_diagram} is an average over about 60 to 85 sets of the 
critical combinations, which belong to different rule tables.

\section{Discussions and Conclusions}


$\odot$  Difference of Patterns with $T(0,0,0,0,0)=h$ and
$T(0,0,0,0,0)=0$ \\

\begin{figure}
\begin{center}
\scalebox{0.55}{ { \includegraphics{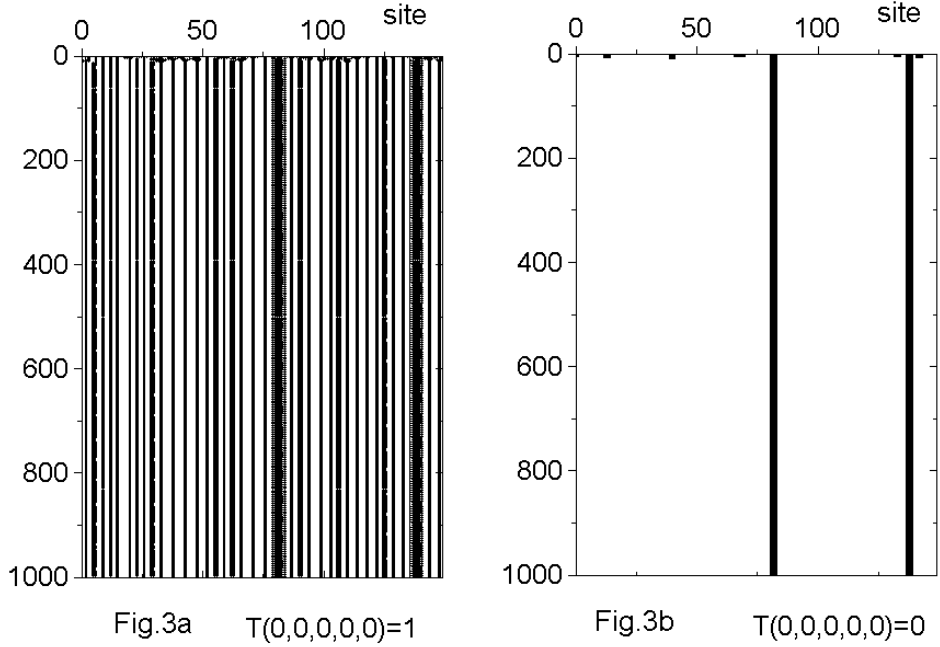} } }
\caption{The difference of the patterns with and without the rule
$T(0,0,0,0,0)=0$ at $\lambda=0.4$.
The rule $T(0,0,0,0,0)=1$ in the Fig.3a is replaced by
$T(0,0,0,0,0)=0$ in the Fig.3b;the RP5 
is carried out.
\label{periopattern} }
\end{center}
\end{figure}

By carrying out the replacements RP4s and RP3s, we could 
artificially obtain the rule table of periodic pattern with
$T(0,0,0,0,0)=h$.
In this case the length of the schema of quiescent state could not
develop longer than 4, then the patterns will have thin stripes in the time
direction. While for the patterns generated by the rule table 
without D5,
there is very small probability that the  thin stripes are created. 
Then there will be clear difference
between the two patterns. An example is shown in the
Fig.\ref{periopattern}.
The differences in patterns is again dramatical while those in rule
table are only 2.

Similarly for the patterns 
of an edge of the chaos, when the rule tables have D5, the
patterns are drawn on the background of the thin stripes as shown in the 
Fig.\ref{periopattern}a,
while if they do not include D5, patterns are drawn essentially on the
quiescent states. The differences between these two patterns are 
impressive.  
Similar situation is observed for the pattern when six D4s are included in 
the rule table. In this case, the maximum length of the schema of quiescent 
state is 3, that the width of the stripe becomes more narrow.
The existence of the rules of the table \ref{destruction}
has  strong effects for the details of the patterns. 

$\odot$ Fluctuation of Rule Tables around $\lambda \sim 0.5$\\

When the rule tables are generated randomly, how many type 1 rules are
included, is determined randomly by the probability $\lambda$.
The largest fluctuation is realized for D5, because it has only one
element. Next is D4 and so on.  And the transition is taken place
about $1.14$ to $1.2$ of the force expressed in the  unit of
$c_5$.
Then the change of the pattern classes are easily realized by the 
fluctuation of $N_5$ and $N_4$ at fixed $\lambda$. This
fact explains why the transition and edge of chaos is scattered rather
wide range in $\lambda$.

Also the fluctuation is largest at $\lambda=0.5$. 
Then the coexistence of the periodic pattern and chaotic
pattern is most likely to be realized around this region.
As the edge of chaos is located at the boundary of these two pattern 
classes, it is also realized most frequently around $\lambda
\sim 0.5$. Especially when the number of the states in cell is 
equal to two.

$\odot$ Edge of Chaos and Order of Transition\\

When the replacements of RP3 or RP2 
are carried out,
we find the edge of chaos(very long transient lengths) in many cases.  
Sometimes they are
observed rather wide range in $N_3$ or $N_2$.
This seems to indicate that in 
many cases, the transition is second order like. But the widths in the
ranges of $N_3$ or $N_2$ are different from each other and there are cases
where the widths
are less than one unit in the replacement of D2 rules(first order like). 
What is the origin of the difference in the width is an open problem and
may be studied by taking into account the effects of the type 3 
rules.  It is a very interesting problem under what condition,
the transition becomes first order like
or second order like.

$\odot$ Special Cases of the  Rule Table\\

The relative strength of the
force given by the table \ref{relative_force}
could not be applied for some special cases.
When the rule table contains all the 33 D3 rules, the
schemas of the quiescent state with length 4 could not develop, then
the presence of the D4 have no chance to work. In this case, the force 
from the D4 and D5 are equal to zero.  The results given in the table
\ref{relative_force} do not include in these special cases. But
these special rule
tables are very hard to be realized as far as they are determined
randomly. The results for the table \ref{relative_force} are 
the average over the
rule tables which are typically obtained randomly according to the
probability $\lambda$.

We have made a few replacements experiments at more extreme $\lambda$.
At $\lambda=0.75$, in most cases, a randomly generated rule tables
belong to chaotic pattern. We have succeeded to change them to
the rule table of periodic pattern by the replacements of RP5, RP4s
and RP3s.
On the other hand at $\lambda=0.2$, most of the generated rule
tables were periodic ones. 
In this case too, the
replacements RC5 and a few RC4s make the rule table to that of 
chaotic pattern classes.
Then the conclusion that the number and combination of the type 1 rule 
determine the pattern classes is correct for more wider
space of rule tables. 
However when $\lambda$ exceed $0.75$, the state $0$ lost its role as a
quiescent state.  Then the
analysis in this article will not be applied there.

$\odot$ Other One Dimensional Cellular Automata\\

The similar study has been carried out for the one dimensional 
3-neighbors and 16-states CA.
In this case there are 3 groups of rules which destroy the schemas of 
quiescent
state; $T(0,0,0)$ and  $T(0,0,i)$, $T(i,0,0)$ and $T(i,0,j)$. 
The results are very preliminary, but we have succeeded to change the
chaotic patterns to periodic one by making only one set of replacements
at $\lambda=\frac{2648}{16^3}$.
\begin{equation}
\begin{array}{ll}
T(0,0,0)=h \rightarrow T(0,0,0)=0 \\
T(\mu,k,\nu)=0 \rightarrow T(\mu,k,\nu)=l
\end{array}
\end{equation}
It needs some more studies to obtain the quantitative results.
But the result that the rules which breaks the schemas of quiescent
state play main role for the determination of the pattern classes 
and the qualitative phase diagram 
will be common for all CA including two dimensional ones.

$\odot$ Conclusions\\

The pattern in the CA is strongly affected by the type 1 rules,
which are classified in the table \ref{destruction}.
Around  $0.4 < \lambda< 0.6$ region, the replacements of D5 and 
a few D4s change the pattern classes. The number of the 
replacements has been 
always less than $1\%$ if RP5 and RP4 or RC5 and RC4 are included. 

The type 1 rules have a hierarchy in their effect to push
the rule table toward  chaotic one as shown in Eq. \ref{order_force}. 
This property is studied more quantitatively by introducing force for 
the rules in table \ref{destruction}. Then we have obtained the phase
diagram of CA. It is determined by many approximations and assumptions.
But it tell us how to control the pattern classes at fixed $\lambda$. 

The edge of the chaos of the
CA is realized as the result of the 
delicate balance 
between the creation of the schemas of quiescent state by type 3 rules
and the destruction of them by the type 1 rules.  By making
replacements 
defined by the Eq.\ref{toperio} or Eq. \ref{tochaos}, 
we could control the pattern classes more efficiently than
using only the parameter $\lambda$.

The results obtained in this article is a beginning of the more
quantitative study of the rule tables and pattern classes.  
The analysis of the structure of the rule table including the effects
of the type 3 rules will be necessary to understand the order of the
transition and the detailed properties of the edge of chaos, which will
be reported elsewhere.

\end{document}